\documentstyle[preprint,aps]{revtex}
\renewcommand{\theequation}{\arabic{section}.\arabic{equation}}
\tighten
\begin{document}
\draft 
\title{The analytic structure of heavy quark propagators} 
\author{C.\ J.\ Burden \vspace*{0.2\baselineskip}}
\address{
Department of Theoretical Physics,
Research School of Physical Sciences and Engineering,
Australian National University, Canberra, ACT 0200, Australia
\vspace*{0.2\baselineskip}\\}
\maketitle
\begin{abstract}
The renormalised quark Dyson-Schwinger equation is studied in the limit 
of the renormalised current heavy quark mass $m_R\rightarrow \infty$.  
We are particularly interested in the analytic pole structure of the 
heavy quark propagator in the complex momentum plane.  Approximations 
in which the quark-gluon vertex is modelled by either the bare vertex or 
the Ball-Chiu Ansatz, and the Landau gauge gluon propagator takes either 
a gaussian form or a gaussian form with an ultraviolet asymptotic tail 
are used.  
\end{abstract}
\section{Introduction}

The solution of approximate Dyson-Schwinger equations (DSE) has 
proved to be an effective means for modelling quark propagators in 
hadronic physics~\cite{RW94}.  Recent calculations within the genre of models 
which we shall refer to as the DSE technique include those of the light 
hadron spectrum~\cite{BSpapers,JM93} and of electromagnetic form 
factors of the pion and kaon~\cite{R96}.  Although no rigorous proof 
exists, it is the philosophy of the DSE technique that one possible 
signal of confinement in QCD should be the absence of timelike poles 
in the quark propagator~\cite{RWK92}.  It has furthermore been conjectured 
that the propagator $S(p)$ could be an entire function in the complex 
$p^2$ plane~\cite{M86,BRW92}.  Such a scenario would, for instance, avoid 
certain unpleasant consequences which can result when modelling mesons 
via the Bethe-Salpeter equation which samples the quark propagator over a 
region of the complex plane.  

Determining the analytic structure of fermion propagators in 
QCD~\cite{MH92,SC92} or other confining theories~\cite{M94,AB96} by the 
direct solution of model DSEs is not easy.  It appears that the pole 
or branch cut structure obtained in any particular model 
is heavily dependent on the approximations employed.  In general, 
two aspects of the quark DSE must be approximated: the quark-gluon 
vertex, and the gluon propagator.  In this paper, we shall look at both 
these aspects within the heavy quark sector.  
In existing numerical studies in the light quark sector which produce 
propagators with conjugate 
singularities~\cite{MH92,SC92}, the quark gluon vertex has usually been 
approximated by the bare vertex (the so called rainbow approximation).  
In ref.~\cite{BRW92}, however, it was shown that an entire function propagator 
can be obtained if the vertex function is modelled by a more sophisticated 
form respecting the Ward-Takahashi identity.  This suggests that 
it is worthwhile exploring the importance of accurately modelling the 
quark-gluon vertex, as well as the gluon propagator, when studying the 
analytic structure of the quark propagator.  

In a recent development, the DSE technique has been extended 
to the realm of heavy quarks~\cite{BL97,DL97} in a way 
inspired by heavy quark effective theory (HQET)~\cite{N94}.  
The purpose of this exercise was twofold.  Firstly, if one acknowledges 
the success of the DSE technique in the light quark sector, it is clear 
that the dynamics of confined particles is driven by non-perturbative 
dynamical self dressing.  In HQET, non-perturbative self dressing 
and the detailed analytic structure of the heavy quark propagator are 
largely ignored.  It is important to know whether this is justified, or 
whether the successes of HQET are purely fortuitous.  Secondly, one has 
the hope that an accurate determination of the heavy quark propagator 
will eventually prove useful for building phenomenological models of 
heavy quark hadrons.  

In ref.~\cite{BL97} a preliminary attempt is made to calculate the 
spectrum of heavy quark--light antiquark mesons by using the combination 
of rainbow DSE and ladder Bethe-Salpeter equation (BSE).  It is found 
that, within the limitations of the model, the pole structure of the 
heavy quark propagator prevents solution of the meson BSE.  This is 
clearly a shortcoming of the approximations involved.  In 
ref.~\cite{DL97} the heavy quark DSE is examined from the point of 
view of the gauge technique.  This is essentially an improvement on 
the rainbow approximation to the quark-gluon vertex which is designed 
to respect the Ward-Takahashi identity.  An alternative approach, and one 
which we follow in this paper, is to replace the bare vertex Ansatz with 
the Ball-Chiu vertex Ansatz~\cite{BC80}.  We shall see that in the heavy 
fermion limit the gauge technique and the Ball-Chiu vertex are equivalent.  

Regarding the gluon propagator, our treatment differs from ref.~\cite{DL97} 
in that it is principally numerical, enabling us to concentrate on a 
more realistic class of model gluon propagators.  Specifically we study 
the simple gaussian Ansatz employed in ref.~\cite{BL97}, designed to 
model the infrared enhanced behaviour of the gluon propagator, and a model 
proposed by Frank and Roberts (FR)~\cite{FR96} which includes both an 
enhanced infrared behaviour and the known asymptotic ultraviolet behaviour.  
In order to deal with the FR propagator, it has been necessary to formulate a 
properly renormalised version of the heavy quark DSE of ref.~\cite{BL97}.  

In summary, our main finding is as follows: Improving the vertex Ansatz 
does little to improve the analytic structure of the quark propagator.  
However, improving the Ansatz employed for the gluon propagator, particularly 
by including a realistic asymptotic ultraviolet tail, moves poles in the 
heavy quark propagator to a less intrusive part of the complex momentum 
plane.  This portends well for future application of the heavy quark 
DSE technique.  

The layout of the paper is as follows.  In Section II we summarise the 
renormalised quark DSE and the approximations we shall be employing for the 
quark-gluon vertex and gluon propagator.  In Section III we 
summarise the heavy quark formalism and derive a renormalised DSE for 
the heavy quark propagator.  Numerical solutions to these equations are 
discussed in Section IV. Conclusions are drawn and suggestions for the 
direction of future work are given in Section V.  

\section{The quark Dyson-Schwinger equation}

Our starting point is the renormalised quark DSE~\cite{RW94} 
\begin{equation}
\Sigma'(p,\Lambda) = Z_1(\mu^2,\Lambda^2)\frac{4g^2}{3} 
 \int^\Lambda \frac{d^4q}{(2\pi)^4}\, 
D_{\mu\nu}(p - q) \gamma_\mu S(q) \Gamma_\nu(q,p),   \label{DSEqn}
\end{equation}
where we have used a Euclidean metric in which timelike vectors satisfy 
$p^2 = -p_{\rm Minkowski}^2 < 0$, 
and for which $\{\gamma_\mu,\gamma_\nu\} = 2\delta_{\mu \nu}$.  
Our aim is to solve the DSE for the renormalised quark propagator 
$S(p,\mu)$, which we write in the form
\begin{eqnarray}
S(p,\mu) & = & \frac{1}{i\gamma\cdot p A(p^2,\mu^2) + B(p^2,\mu^2)} 
                                       \nonumber \\
    & = & \frac{1}{Z_2(\mu^2,\Lambda^2)[i\gamma\cdot p + m_0(\Lambda)] 
               + \Sigma^{'}(p,\Lambda)} .   \label{genprop}
\end{eqnarray}
The unrenormalised self energy is written
\begin{equation} 
\Sigma'(p,\Lambda) = i\gamma\cdot p \left[A'(p^2,\Lambda^2) - 1\right]
       + B'(p^2,\Lambda^2).  \label{Sigpr}
\end{equation}

If the renormalisation scale is set such that 
\begin{equation}
\left.S(p)\right|_{p^2 = \mu^2}=\frac{1}{i\gamma\cdot p + m_R(\mu^2)},
                            \label{mufix}
\end{equation}
it follows from Eqs.~(\ref{genprop}) and (\ref{Sigpr}) that renormalised 
and bare quantities are related by 
\begin{equation}
 Z_2(\mu^2,\Lambda^2) = 2 - A'(\mu^2,\Lambda^2),  \label{Zfix}
\end{equation}
\begin{equation}
 A(p^2,\mu^2) = 1 + A'(p^2,\Lambda^2) - A'(\mu^2,\Lambda^2), \label{AApr}
\end{equation}
\begin{equation}
 B(p^2,\mu^2) = m_R(\mu^2) + B'(p^2,\Lambda^2) - B'(\mu^2,\Lambda^2). 
                     \label{BBpr}
\end{equation}

The set of equations (\ref{DSEqn}) and (\ref{Zfix}) to 
(\ref{BBpr}) together with the `abelian approximation' $Z_1 = Z_2$ can 
be solved numerically for the propagator functions $A$ and $B$ once the 
renormalised quark-gluon vertex function $\Gamma_\mu$, the renormalised 
gluon propagator $D_{\mu \nu}$, the renormalisation point 
$\left(\mu,m_R(\mu)\right)$ and cutoff $\Lambda$ are specified.  
The precise forms 
of the quark-gluon vertex and gluon propagator are unknown and must be 
modelled by appropriate Ans\"{a}tze.  We next summarise the Ans\"{a}tze 
employed in this paper.  

\subsection{Quark-gluon vertex}

The most general form of the quark-gluon vertex consistent with Lorentz and 
CPT invariance, satisfying the Ward identity 
$i\Gamma_\mu(p,p) = \partial^p_\mu S^{-1}(p)$ and Ward Takahashi 
identity\footnote{Strictly speaking, it is the Slavnov-Taylor identities, 
 which include ghost contributions, and not the Ward-Takahashi identities 
 which are relevant to QCD.  By using the Ball-Chiu vertex we are 
 effectively ignoring the ghost self energy and ghost-quark scattering 
 kernel.  This is a commonly used approximation~\cite{RW94}.}  
$i(p - q)_\mu \Gamma_\mu(p,q) = S^{-1}(p) - S^{-1}(q)$, and free of 
kinematic singularities has been given by Ball and Chiu~\cite{BC80}.  
It takes the form 
\begin{equation}
\Gamma_\mu(p,q) = \Gamma_\mu^{\rm BC}(p,q) + \Gamma_\mu^{\rm T}(p,q) , 
                                        \label{genver}
\end{equation}
where 
\begin{eqnarray}
\Gamma_\mu^{\rm BC}(p,q) & = & \frac{1}{2} \left[A(p^2) + A(q^2)\right]
                       \gamma_\mu           \nonumber \\
& + & \frac{(p + q)_\mu}{(p^2 - q^2)}\left\{ \left[A(p^2) - A(q^2)\right]
       \frac{\gamma\cdot p + \gamma\cdot q}{2} - 
          i \left[B(p^2) - B(q^2)\right] \right\},  \label{BCver}
\end{eqnarray}
and $\Gamma_\mu^{\rm T}(p,q)$ is an otherwise unconstrained piece 
satisfying the conditions $(p - q)_\mu \Gamma_\mu^{\rm T}(p,q) = 0$ and 
$\Gamma_\mu^{\rm T}(p,p) = 0$.  

We mention two well studied vertex Ans\"{a}tze falling within this class.  
The first of these, introduced 
by Curtis and Pennington~\cite{CP90} to ensure multiplicative 
renormalisability in quantum electrodynamics, is defined by setting the 
transverse piece $\Gamma_\mu^{\rm T}$ equal to 
\begin{equation}
\Gamma_\mu^{{\rm T}_{\rm CP}}(p,q) = \frac{A(p^2) - A(q^2)}{2d(p,q)} 
    \left[\gamma_\mu(p^2 - q^2) - (p + q)_\mu 
              (\gamma\cdot p - \gamma\cdot q ) \right], \label{CPTr}
\end{equation}
with
\begin{equation}
d(p,q) = \frac{(p^2 - q^2)^2 + \left[ M^2(p^2) + M^2(q^2)^2 \right]^2}
               {p^2 + q^2} , 
\end{equation}
where $M = B/A$.  

The second of these, proposed by Haeri~\cite{H91}, takes the form 
\begin{eqnarray}
\Gamma_\mu^{\rm H}(p,q) & = & \frac{p^2 A(p^2) - q^2 A(q^2)}{p^2 - q^2} 
                \gamma_\mu +  \frac{A(p^2) - A(q^2)}{p^2 - q^2} 
           \gamma\cdot p \gamma_\mu \gamma\cdot q \nonumber \\
 & & -i \frac{B(p^2) - B(q^2)}{p^2 - q^2} 
     \left(\gamma\cdot p \gamma_\mu +  \gamma_\mu \gamma\cdot q\right).  
                         \label{Hver}
\end{eqnarray}
It satisfies the above criteria and therefore must be of the form 
Eq.~(\ref{genver}).  Munczek~\cite{Mpr} has shown that the Haeri vertex is 
identical to the spectral representation of the vertex used in the gauge 
technique, and this has in turn been employed in ref.~\cite{DL97}.  

\subsection{Gluon propagator}

In a general covariant gauge, the gluon propagator takes the form 
\begin{equation}
g^2 D_{\mu \nu}(k) = 
  \left(\delta_{\mu \nu} - \frac{k_\mu k_\nu}{k^2}\right) \Delta(k^2) 
        + g^2 \xi \frac{k_\mu k_\nu}{k^4},    \label{gprop}
\end{equation}
where $\xi$ is the gauge fixing parameter.  Perhaps the simplest Ansatz 
which has proved useful for modelling QCD is the `infrared dominant' 
model~\cite{M86,BRW92} 
\begin{equation}
\Delta_{\rm IR}(k^2) = \frac{3}{16}(2\pi)^4 \mu^2 \delta^4(k).  \label{delpr}
\end{equation}
In applications to hadronic physics, 
$\mu$ is usually taken to be of the order of 1 GeV~\cite{MN83}, 
which is the typical scale of QCD.  An obvious disadvantage of this 
model is that it neglects completely any intermediate or ultraviolet 
behaviour.   One can go some way towards incorporating some intermediate 
energy structure by using the computationally convenient gaussian model 
\begin{equation}
\Delta_{\rm G}(k^2) = \frac{3}{16}(2\pi)^4 \frac{\mu^2}{\alpha^2 \pi^2}
                  e^{-k^2/\alpha}. \label{gauss} 
\end{equation}
We note that the infrared dominant gluon model $\Delta_{\rm IR}$, together 
with the minimal Ball-Chiu vertex 
$\Gamma_{\mu \nu}(p,q) = \Gamma_{\mu \nu}^{\rm BC}(p,q)$ defines precisely 
the light quark model considered in ref.~\cite{BRW92}, whereas the 
gaussian model $\Delta_{\rm G}$ has been used in our earlier rainbow 
approximation studies of heavy quarks~\cite{BL97}.  Below we shall 
explore the effect on the analytic structure of the heavy quark propagator 
of combining either $\Delta_{\rm IR}$ or $\Delta_{\rm G}$ with the Ball-Chiu 
vertex.  

A more realistic model gluon propagator which goes some way toward 
modelling the asymptotically free ultraviolet behaviour of QCD 
(neglecting logarithmic corrections) has been proposed by Frank and 
Roberts~\cite{FR96}.  It takes the form 
\begin{equation}
\Delta_{\rm FR}(k^2) = 4\pi^2 d \left[4\pi^2m_t^2\delta^4(k) + 
         \frac{1 - e^{k^2/(4m_t^2)}}{k^2} \right], \label{FRprop} 
\end{equation} 
where $d = 12/(33 - 2N_f)$, $N_f = 3$ is the number of light quark 
flavours, and $m_t \approx 0.69$ GeV is a parameter fitted to a range 
of calculated pion observables.  In our numerical calculations, we find it
more convenient to consider a gaussian smeared version of the FR propagator 
given by 
\begin{equation}
\Delta_{\rm GFR}(k^2) = (2\pi)^4 \frac{m_t^2 d}{\alpha^2 \pi^2}
      e^{-k^2/\alpha} + 4\pi^2 d \frac{1 - e^{k^2/(4m_t^2)}}{k^2}.  
                                          \label{GFRprop}
\end{equation}
This will enable comparisons to be made with the pure gaussian model 
$\Delta_G$.  

In the next section we consider the heavy quark limit of the quark 
Dyson-Schwinger equation.  Our earlier analysis of this limit ~\cite{BL97} 
employed the  heavily damped gluon propagator $\Delta_{\rm G}$, and so 
was free from ultraviolet divergences.  In order to deal with a more 
realistic model, such as Eq.~(\ref{GFRprop}), it is necessary to develop a 
properly renormalised version of the formalism.  

\section{The heavy quark limit} 

\subsection{Heavy quark propagator}

In the absolute limit of heavy renormalised quark masses, 
$m_R(\mu^2) \rightarrow \infty$, 
the dressed quark propagator Eq.~(\ref{genprop}) is dominated by the 
bare form $S^{-1}_{\rm bare} = i\gamma\cdot p + m_R$.  However it is 
important to isolate from the full inverse propagator finite order self energy 
corrections to the bare inverse propagator which drive confining and remnant 
chiral symmetry breaking effects.  To this end we write the momentum 
variables occurring in the DSE Eq.~(\ref{DSEqn}) as 
\begin{equation}
p_\mu = im_R(\mu^2) v_\mu + k_\mu, \hspace{5 mm} 
q_\mu = im_R(\mu^2) v_\mu + k'_\mu,                \label{kmudef}
\end{equation}
where, for convenience, we take $v = ({\bf 0},1)$, so $k\cdot v = k_4$.  
We then write the renormalised quark propagator functions as 
\begin{equation}
A(p^2,\mu^2) = 1 + \frac{\Sigma_A(K,\kappa)}{m_R(\mu^2)},   \label{sigAdef}
\end{equation}
\begin{equation}
B(p^2,\mu^2) = m_R(\mu^2) + \Sigma_B(K,\kappa),  \label{sigBdef}
\end{equation}
where we have defined the independent momentum variable 
\begin{equation}
K = \frac{p^2 + m_R^2}{2im_R} = k_4 + \frac{k^2}{2im_R}, \label{Kdef}
\end{equation}
and renormalisation point 
\begin{equation}
\kappa = \frac{\mu^2 + m_R^2}{2im_R},
\end{equation}
in the complex $K$ plane.  The change of dependent variable 
$p^2 \rightarrow K$ induced by the transformation Eq.~(\ref{Kdef}) 
is illustrated in Fig.~\ref{fig1}. In general, when working to zeroth order 
in $1/m_R$, one can use the approximation $K = k_4$ (an approximation 
which was used in ref.~\cite{BL97}).  An exception to this rule, relevant to  
models such as Eq.~(\ref{GFRprop}) for which the gluon propagator is 
not heavily ultraviolet damped, is in the denominator of the heavy 
quark propagator in the integrand in the quark DSE.  This point will 
become clearer at the end of Section IIIB.  

From Eqs.~(\ref{genprop}), (\ref{sigAdef}) and (\ref{sigBdef}) we have 
\begin{equation}
S(p,\mu) = \frac{1 + \gamma_4}{2} \frac{1}{iK + \Sigma(K,\kappa)} 
    + O\left(\frac{1}{m_R}\right) ,     \label{hprop}
\end{equation}
where we have defined the heavy quark self energy 
\begin{equation} 
 \Sigma(K,\kappa) =  \Sigma_B(K,\kappa) -  \Sigma_A(K,\kappa).\label{Sigdef}  
\end{equation}
We find in general that the DSE leads to a single integral equation for 
the complex valued function $\Sigma(K,\kappa)$.  
The form given by Eq.(\ref{hprop}) represents the heavy quark 
propagator in the dominant region near the the bare propagator mass pole 
$p^2 = -m_R^2$.  Obtaining an integral equation for $\Sigma(K)$ involves 
the change of integration $\int d^4q \rightarrow \int d^4k' = 
\int_{-\infty}^\infty dk'_4 \int_0^\infty d\left|{\bf k}\right|
\left|{\bf k}\right|^2$ induced by the change of variable Eq.~(\ref{kmudef}).  
For this change of integration to be valid, the propagator, and hence 
the functions 
$A(q^2)/[q^2A(q^2)^2 + B(q^2)^2]$ and $B(q^2)/[q^2A(q^2)^2 + B(q^2)^2]$,
must be analytic over the shaded region in Fig.~\ref{fig1}, 
\begin{equation}
{\rm Re}\,(q^2) > -m_R^2 + \frac{\left({\rm Im}\,(q^2)\right)^2}{4m_R^2}.  
\end{equation}
Equivalently, the function defined by 
\begin{equation}
\sigma_{\rm Q}(K,\kappa) = \frac{1}{iK + \Sigma(K,\kappa)}, \label{sgdef}
\end{equation}
must be analytic over the shaded region ${\rm Im}\, K < 0$.  

The confinement criterion that $S(p)$ should be free from timelike poles 
on the negative real $p^2$ axis translates in the heavy quark case 
to a requirement that $\sigma_{\rm Q}$ should be free from poles on 
the imaginary $K$ axis.  The stronger conjecture~\cite{BRW92}, 
that the quark propagator should be an entire function of $p^2$ 
translates in the heavy quark formalism to a conjecture that 
$\sigma_{\rm Q}$ should be an entire function of $K$.  

From Eq.~(\ref{hprop}), we have that the renormalisation condition 
Eq.~(\ref{mufix}) is equivalent to 
\begin{equation} 
\left.S(p,\mu)\right|_{p^2 = \mu^2} = 
  \left. \frac{1 + \gamma_4}{2}\frac{1}{iK} \right|_{K = \kappa}\, ,  
\end{equation}
to zeroth order in $1/m_R$.  Typically we choose $\kappa$ to be on 
the negative imaginary $K$ axis, as the heavy quark propagator asymptotes 
to the bare propagator as $K \rightarrow -i\infty$, as can be seen from 
Fig.~\ref{fig1}.  

\subsection{Heavy quark DSE}

To illustrate the derivation of the renormalised heavy quark DSE, we 
choose Landau gauge ($\xi = 0$ in Eq.~(\ref{gprop})) and, for the time being, 
work with the rainbow or bare vertex approximation
\begin{equation}
\Gamma_\mu(p,q) = \gamma_\mu.  
\end{equation}
Using Dirac trace identities to project out from Eqs.~(\ref{DSEqn}) 
to (\ref{Sigpr}) a pair of coupled integral equations gives 
\begin{eqnarray}
\lefteqn{A'(p^2,\Lambda^2) = }\nonumber \\
& & \!\!\!\!\!\! 1 + \frac{4Z_1}{3p^2}  
  \int^\Lambda \frac{d^4q}{(2\pi)^4}\, \left[p\cdot q + 
  2\frac{p\cdot(p - q) q\cdot(p - q)}{(p - q)^2}\right] 
   \Delta\left[(p - q)^2\right] \frac{A(q^2)}{q^2 A^2 + B^2},  \label{ASDE}
\end{eqnarray}
\begin{equation}
B'(p^2,\Lambda^2) = 4Z_1  \int^\Lambda \frac{d^4q}{(2\pi)^4}\, 
   \Delta\left[(p - q)^2\right] \frac{B(q^2)}{q^2 A^2 + B^2}.  \label{BSDE}
\end{equation}
Substituting 
\begin{equation}
 \frac{A(q^2)}{q^2 A(q^2)^2 + B(q^2)^2} = 
 \frac{1}{2m_R}\,\frac{1}{iK' + \Sigma(K')}  + O\left(\frac{1}{m_R^2}\right) ,
\end{equation}
\begin{equation}
 \frac{B(q^2)}{q^2 A(q^2)^2 + B(q^2)^2} = 
 \frac{1}{2}\,\frac{1}{iK' + \Sigma(K')}  + O\left(\frac{1}{m_R}\right) ,
\end{equation}
into Eqs.~(\ref{ASDE}) and (\ref{BSDE}) gives 
\begin{eqnarray}
m_R\left[A'(p^2,\Lambda^2) - 1\right] & = & 
    \frac{2Z_1}{3} \int^\Lambda 
       \frac{d^4k'}{(2\pi)^4}\,\left[1 + 2\frac{(k_4 - k'_4)^2}{(k - k')^2}
      \right]   \nonumber \\
   & & \times \Delta\left[(k - k')^2\right] \frac{1}{iK'+\Sigma(K',\kappa)}
                    + O\left(\frac{1}{m_R}\right) , \label{AprSDE}
\end{eqnarray}
and 
\begin{equation}
 B'(p^2,\Lambda^2) = 2Z_1  \int^\Lambda \frac{d^4k'}{(2\pi)^4}\,
   \Delta\left[(k - k')^2\right] \frac{1}{iK'+\Sigma(K',\kappa)}
                    + O\left(\frac{1}{m_R}\right) . \label{BprSDE}
\end{equation} 
From Eqs.~(\ref{AApr}), (\ref{BBpr}), (\ref{sigAdef}), (\ref{sigBdef}) 
and (\ref{Sigdef}) we have 
\begin{equation}
\Sigma(K,\kappa) = \left[B'(p^2,\Lambda^2) - m_R A'(p^2,\Lambda^2)\right]
                    -[p^2\rightarrow\mu^2], 
\end{equation}
which gives 
\begin{equation}
\Sigma(K,\kappa) = \frac{4}{3} 
   \int^\Lambda \frac{d^4k'}{(2\pi)^4}\,
 \left[ \frac{\left|{\bf k} - {\bf k}'\right|^2}{(k - k')^2} 
   \Delta\left[(k - k')^2\right] \frac{1}{iK'+\Sigma(K',\kappa)}
           - (K\rightarrow\kappa) \right] , \label{hsde1}
\end{equation} 
where $K$ on the left hand side is related to $k$ under the integrand 
via Eq.~(\ref{Kdef}), and a similar definition exists relating $K'$ and $k'$.  
Here we have assumed that Eqs.~(\ref{Zfix}) and (\ref{AprSDE}) imply 
$Z_1 = 1 + O(1/m_R)$.  

As noted above, $K$ and $k_4$ are interchangeable to leading order, except 
where they occur in the denominator of the heavy fermion propagator 
$1/[iK' + \Sigma(K')]$ in the integrand of Eq.~(\ref{hsde1}).  
This is because, for any gluon propagator $\Delta$ with a realistic 
asymptotic UV behaviour, all powers of $k'$ must be retained in the 
denominator to maintain the same degree of divergence in the heavy 
fermion DSE as in the original equation (\ref{DSEqn}).  With this observation, 
and choosing $k_\mu = ({\bf 0},K)$ and the renormalisation point 
$({\bf 0},\kappa)$, we arrive at the leading order heavy quark DSE 
\begin{eqnarray}
\Sigma(K,\kappa) & = & \frac{4}{3} \int^\Lambda \frac{d^4k'}{(2\pi)^4}\, 
     \frac{1}{ik_4' + k'^2/(2m_R) + \Sigma(k'_4,\kappa)} \nonumber \\
 & & \times \left\{ \frac{\left|{\bf k}'\right|^2 
          \Delta\left[(K - k'_4)^2 + \left|{\bf k}'\right|^2 \right]}
 {(K - k'_4)^2 + \left|{\bf k}'\right|^2}  - (K \rightarrow\kappa) \right\}.
                                  \label{hsde2}
\end{eqnarray}  
We show in the Appendix that, with the smeared FR gluon 
propagator Ansatz Eq.~(\ref{GFRprop}), and assuming a hierarchy of scales
\begin{equation}
m_t,\;\left|K\right|,\;\left|\kappa\right| \ll m_R \ll \Lambda,  \label{hier}
\end{equation}
the integral in Eq.~(\ref{hsde2}) is independent of the ultraviolet cutoff 
$\Lambda$, and the heavy quark self energy behaves like 
\begin{equation}
 \Sigma(K,\kappa) \sim  2id(\kappa - K) \ln\left(\frac{m_R}{m_t}\right), 
\end{equation}
as $m_R\rightarrow\infty$.  With the more severely truncated gaussian 
gluon propagator Eq.~(\ref{gauss}), the renormalisation point $\kappa$ can 
be taken to $-i\infty$ and the ${k'}^2/(2m_R)$ term in the denominator 
of the integrand ignored with impunity.  

\subsection{Choice of renormalisation point}

At the end of the day, physical quantities must be insensitive to the 
choice of renormalisation point $\kappa$.  In this section we note that 
the freedom to choose the renormalisation point is equivalent to the 
notion of a `residual mass' in HQET~\cite{FNL92,N94}, that is, the notion 
that to zeroth order in $1/m_R$, physical quantities computed in HQET do not 
depend on the choice of $m_R$.  

After formally carrying out the spatial momentum integration in 
Eq.~(\ref{hsde2}), one obtains an equation generically of the form 
\begin{equation}
\Sigma(K,\kappa) = \frac{1}{\sigma_Q(K,\kappa)} - iK 
  = \int_{-\infty}^\infty dK' \, \left[T(K - K') - T(\kappa - K')\right]
         \sigma_Q(K',\kappa),   \label{frmly}
\end{equation}
for some kernel $T$, and with $\sigma_Q$ defined by Eq~(\ref{sgdef}).  
It is possible to show from this generic form that the effect of making 
a change of renormalisation point 
$\kappa_{\rm old} \rightarrow \kappa_{\rm new}$ is equivalent to a shift 
of the quark propagator solution along the imaginary $K$ axis: 
\begin{equation}
\sigma_Q(K,\kappa_{\rm new}) = \sigma_Q(K - i\delta m,\kappa_{\rm old}), 
                       \label{shift}
\end{equation}
where $\delta m$ is the solution to 
\begin{equation}
\frac{1}{i \kappa_{\rm new}} = 
  \sigma_Q(\kappa_{\rm new} - i\delta m,\kappa_{\rm old}),  \label{delm}
\end{equation} 
that is, the shift is that required to ensure that the new heavy quark 
propagator pass through the renormalisation point 
$\sigma_Q(\kappa_{\rm new},\kappa_{\rm new}) = 1/i\kappa_{\rm new}$.  
Referring to Fig.~\ref{fig1} we see that this is equivalent to a shift  
in the position of the origin of the $K$ plane along the $p^2$ axis 
corresponding to changing $m_R$ by an amount $\delta m$.  

One can also demonstrate that the mass differences between any two 
heavy quark--light antiquark meson states calculated from the Bethe-Salpeter 
formalism set out in ref.~\cite{BL97} is independent of the renormalisation 
point.  

\section{Results}

Our main concern in this paper is to compare how the analytic structure of 
the heavy quark propagator solutions is affected by the approximations 
employed both for the quark-gluon vertex and the gluon propagator.  In an 
earlier work~\cite{BL97} an attempt was made to study the heavy quark--light 
antiquark meson spectrum using a combination of rainbow DSE and ladder 
BSE.  It was found that, if a simple gaussian Ansatz is used for the gluon 
propagator, complex conjugate poles occur in the heavy quark propagator 
which prevent solution of the meson BSE.  Below we systematically explore 
the movement of the poles as the bare vertex of the rainbow approximation 
is replaced by a Ball-Chiu vertex, and as the gaussian gluon propagator is 
replaced by the more realistic Frank and Roberts propagator.  

\subsection{Gaussian gluon propagator} 

\subsubsection{Rainbow approximation}

As noted in the previous section, if the gluon propagator Eq.~(\ref{gauss}) 
is used, we may set the renormalisation point $\kappa = -i\infty$ 
and ignore the ${k'}^2/(2m_R)$ term in Eq.~(\ref{hsde2}).  Furthermore, 
there is no need to distinguish between the independent momentum variables 
$K$ and $k_4$.  With these simplifications the Landau gauge, rainbow 
heavy quark DSE becomes 
\begin{equation}
\Sigma(k_4) = \frac{4}{3} \int \frac{d^4k'}{(2\pi)^4} \,
    \frac{\left|{\bf k} - {\bf k}'\right|^2}{(k - k')^2} 
      \frac{\Delta\left((k - k')^2\right)}
                {ik_4' + \Sigma(k_4')}.  \label{baresde}
\end{equation}
Choosing the infrared dominant gluon propagator Eq.~(\ref{delpr}), the DSE 
reduces to an algebraic equation with solution 
\begin{equation}
\Sigma(k_4) =
\left \{ \begin{array}{lrl}
\frac{1}{2} \left(-ik_4 + \sqrt{\frac{3\mu^2}{4} - k_4^2} \right)
         & \mbox{if } & 0\le k_4 < \frac{\sqrt{3}\mu}{2} , \\
                                  &            &            \\
\frac{i}{2} \left(- k_4 + \sqrt{k_4^2 - \frac{3\mu^2}{4}} \right)
         & \mbox{if } & k_4\ge \frac{\sqrt{3}\mu}{2} ,
                   \end{array}   \right.   \label{baresoln}
\end{equation}
or, using the definition (\ref{sgdef}), 
\begin{equation}
\sigma_{\rm Q}(k_4) =
\left \{ \begin{array}{lrl}
\frac{8}{3\mu^2} \left(-ik_4 + \sqrt{\frac{3\mu^2}{4} - k_4^2} \right)
         & \mbox{if } & 0\le k_4 < \frac{\sqrt{3}\mu}{2} , \\
                                  &            &            \\
\frac{8i}{3\mu^2} \left(- k_4 + \sqrt{k_4^2 - \frac{3\mu^2}{4}} \right)
         & \mbox{if } & k_4\ge \frac{\sqrt{3}\mu}{2} ,
                   \end{array}   \right.   \label{delsig}
\end{equation}

Alternatively, choosing the gaussian gluon propagator Eq.~(\ref{gauss}) and 
carrying out the $d^3{\bf k}$ integration, we obtain the integral 
equation~\cite{GR80} 
\begin{eqnarray}
\lefteqn{\Sigma(k_4) = \frac{\mu^2}{2\alpha^2\sqrt{\pi}} 
   \int_{-\infty}^\infty dk_4 \, \frac{1}{ik_4' + \Sigma(k_4')} \times}
                  \nonumber \\
& & \left\{\sqrt{\alpha} \left[\frac{\alpha}{2} - (k_4 - k_4')^2 \right] 
      e^{-(k_4 - k_4')^2/\alpha} + \sqrt{\pi} \left|k_4 - k_4'\right|^3 
   {\rm erfc}\left(\frac{\left|k_4 - k_4'\right|}
                     {\sqrt{\alpha}}\right) \right\},  
                            \label{bareie}
\end{eqnarray}
where ${\rm erfc}\,z = 1 - {\rm erf}\,z$ is the complementary error function.  
This equation can be solved numerically.  

\subsubsection{Ball-Chiu vertex}

If any of the minimal Ball-Chiu Ansatz Eq.~(\ref{BCver}), the 
Curtis-Pennington Ansatz Eqs.(\ref{genver}) and (\ref{CPTr}), or the 
Haeri Ansatz Eq.~(\ref{Hver}) is used in place of the bare vertex, together 
with the Landau gauge gluon propagator, we obtain in place of 
Eq.~(\ref{baresde}) the equation
\begin{equation}
\Sigma(k_4) = \frac{4}{3} \int \frac{d^4k'}{(2\pi)^4} \,
    \frac{\left|{\bf k} - {\bf k}'\right|^2}{(k - k')^2} 
      \frac{\Delta\left((k - k')^2\right)}{ik_4' + \Sigma(k_4')}
         \left[1 + \frac{\Sigma(k_4) - \Sigma(k'_4)}{i(k_4 - k'_4)}\right].  
                                \label{bcsde}
\end{equation}
It is interesting to note that, within the set of vertex Ans\"{a}tze 
we have considered, the heavy quark propagator is insensitive to the 
transverse part of the vertex.  This is not difficult to understand for the 
Curtis-Pennington vertex, in which the transverse part is heavily 
damped by the presence of the factor $M^4 \sim m^4$ in the denominator 
$d(p,q)$.  However, in the case of the Haeri vertex there is no such 
obvious mechanism, and one is led to question whether the heavy quark 
propagator may be insensitive to a broad class of Ans\"{a}tze satisfying 
the criteria specified above Eq.~(\ref{genver}).  

Taking the gluon propagator to be the infrared dominant form 
Eq.~(\ref{delpr}), gives the differential equation 
\begin{equation} 
\Sigma(k_4) = \frac{3\mu^2}{16i} \frac{d}{dk_4}\ln[ik_4 + \Sigma(k_4)], 
\end{equation}
which, together with the boundary condition $\sigma_{\rm Q}(k_4) 
\rightarrow 0$ as $k_4 \rightarrow -i\infty$, admits the solution 
\begin{equation}
\sigma_{\rm Q}(k_4) = \beta\left[\sqrt{\pi} e^{-\beta^2k_4^2} 
              + 2iF(-\beta k_4) \right],  \label{bcsoln}
\end{equation}
where $\sigma_Q$ is defined by Eq.~(\ref{sgdef}), 
$\beta = 2\sqrt{2}/\mu\sqrt{3}$ and 
\begin{equation}
F(z) = e^{-z^2}\int_0^z e^{t^2} dt = \frac{i\sqrt{\pi}}{2} e^{-z^2}
             {\rm erf}(-iz), 
\end{equation}
is Dawson's integral.  
We note that this solution is an entire function of $k_4$, which, as 
pointed out earlier, is a desirable feature of a quark propagator.  This 
comes as no surprise, as it simply the heavy quark limit of the model 
considered in ref.~\cite{BRW92}, in which it was demonstrated that 
the combination of Ball-Chiu vertex and infrared dominant gluon 
propagator leads to an entire function propagator for all values of 
the bare current quark mass.  

It is of interest to determine to what extent this analytic structure 
is a feature of the Ball-Chiu vertex, and to what extent it is a feature 
of the infrared dominant gluon propagator.  If the infrared dominant 
propagator is replaced by the gaussian smeared form Eq.~(\ref{gauss}), we 
obtain the integral equation 
\begin{eqnarray}
\lefteqn{\Sigma(k_4) = \frac{\mu^2}{2\alpha^2\sqrt{\pi}} 
   \int_{-\infty}^\infty dk_4 \, \frac{1}{ik_4' + \Sigma(k_4')} \times}
                  \nonumber \\
& & \left\{\sqrt{\alpha} \left[\frac{\alpha}{2} - (k_4 - k_4')^2 \right] 
      e^{-(k_4 - k_4')^2/\alpha} + \sqrt{\pi} \left|k_4 - k_4'\right|^3 
   {\rm erfc}\left(\frac{\left|k_4 - k_4'\right|}
                     {\sqrt{\alpha}}\right) \right\}
                  \nonumber \\
& & \times   
     \left[1 + \frac{\Sigma(k_4) - \Sigma(k'_4)}{i(k_4 - k'_4)}\right],  
                    \label{bcie}
\end{eqnarray}
which can be solved numerically.  

\subsubsection{Numerical results: gaussian gluon propagator}

For the purpose of determining the analytic structure of the heavy 
fermion propagator obtained from the DSE with the gaussian gluon propagator 
Eq.~(\ref{gauss}), it is sufficient to look at the one parameter family 
of models obtained by scaling either $\mu$ or $\alpha$ to unity.  We choose 
to scale $\mu$ to unity, which amounts to working with a set of 
dimensionless quantities
\begin{equation}
  \hat{k}_4 = k_4/\mu, \hspace{5 mm} \hat{\alpha} = \alpha/\mu^2, 
  \hspace{5 mm} \hat{\sigma}_{\rm Q} = \mu\sigma_{\rm Q}.  
\end{equation} 
This choice enables us to recover the infrared dominant model in the 
limit $\alpha \rightarrow 0$.  

In Figs.~\ref{fig2} and \ref{fig3} we plot the heavy quark self energy 
$\Sigma(k_4)$ as a function of real $k_4$ obtained from the bare vertex 
DSE Eq.~(\ref{bareie}) and the Ball-Chiu vertex DSE Eq.~(\ref{bcie}) 
for $\hat{\alpha} = 1$, 2 and 3.  These results are obtained by iterating 
from an initial guess and using a Simpson's rule quadrature.  We find 
that the derivative-like terms in Eq.~(\ref{bcie}) prevent a numerical 
solution for values of $\hat{\alpha}$ less than 1, as numerical 
noise in the function values 
becomes unstable with respect to iteration at small values of $k_4$.  
This problem is a general feature of numerical treatments of DSEs with 
Ball-Chiu-like vertices.  Also plotted are the $\hat{\alpha} = 0$ analytic 
results Eqs.~(\ref{baresoln}) and (\ref{bcsoln}).  In all cases the 
self energy is characterised by a real part which peaks at zero and an 
imaginary part which peaks near the typical scale of the model 
$k_4 \sim \mu$.  The self energy for negative real $k_4$ can be obtained 
from these results using the reflection property $\Sigma(-k_4^*) = 
\Sigma(k_4)^*$.  

To solve for the heavy quark propagator away from the real $k_4$ axis 
we shift the contour of integration into the complex plane parallel to 
the real $k_4$ axis and again solve iteratively.  We note that, to 
determine $\Sigma(k_4)$ for complex arguments, it is 
necessary to move the contour of integration to pass through the point 
$k_4$.  This is because the radial part of the $d^3{\bf k}'$ 
integration, carried out in going from Eq.~(\ref{baresde}) to 
Eq.~(\ref{bareie}) or from Eq.~(\ref{bcsde}) to Eq.~(\ref{bcie}), 
creates a pinch singularity at $k'_4 = k_4$ in the error function term 
in Eq.~(\ref{bareie}) or Eq.~(\ref{bcie}).  

We have carried out a search for poles in the propagator function 
$\sigma_Q(k_4)$ for a range of values of $\alpha$ for both the bare and 
Ball-Chiu vertex. Our results are listed in Table~\ref{tab}.  In all cases 
we find that the only observed poles occur for ${\rm Im}\, k_4 > 0$, 
and that $\sigma_Q$ dies away to small values and is free from singularities 
over that part of the shaded region in Fig.~\ref{fig1} accessible to our 
computer program.  Of course we are unable to pass the 
contour of numerical integration through the pole itself, 
and these results are attained by extrapolation from results of contours 
which we gradually moved deeper into the complex plane.  We were unable 
to obtain a reasonable extrapolation for the Ball-Chiu vertex at $\alpha = 1$, 
again because of the iterative instability problem associated with 
the derivative-like term in Eq.~(\ref{bcie}).  

We also list in Table~\ref{tab} the results of using the model gluon 
propagator 
\begin{equation}
g^2 D_{\mu \nu}(k^2) = 
  \delta_{\mu \nu} \Delta_{\rm G}(k^2) ,    \label{flg}
\end{equation}
where $\Delta_{\rm G}$ is given by Eq.~(\ref{gauss}).  
Forms such as Eq.~(\ref{flg}) are frequently used in phenomenological 
modelling (see for instance ref.~\cite{BSpapers}) 
and are sometimes referred to as propagators in a `Feynman-like gauge', 
though of course they are generally not of the form of Eq.~(\ref{gprop}).  
The sole advantage of the Feynman-like gauge is that it leads to considerably 
simplified calculations. In our case it is possible 
to locate poles more accurately because there is no pinch singularity 
requiring the contour of integration to pass through the point in question.  
Once the propagator has been solved on the real $k_4$ axis, the value of 
the propagator can be calculated at any point in the complex plane by 
integrating once along the real axis.  Nevertheless, we have also 
repeated our pole calculations by shifting the contour and extrapolating 
as in the Landau gauge case as a check on the the consistency of the 
two methods and find that they agree to within the accuracy given in 
Table~\ref{tab} of the corresponding Landau gauge results.  
A Feynman-like gauge propagator was also used in 
ref.~\cite{BL97} dealing with the ladder Bethe-Salpeter equation for 
the heavy quark--light antiquark system.  There it was demonstrated that 
the model with bare quark gluon  vertex and gaussian Feynman-like gauge 
gluon propagator had no solutions because of poles in the heavy and 
light quark propagators.  Ideally, one would like improvements in the 
DSE approximations to move the poles further from the real $k_4$ axis 
to avoid the region of the $k_4$ plane sampled by a Bethe-Salpeter 
calculation.  

From Table~\ref{tab} we conclude that simply replacing the bare vertex 
by the Ball-Chiu vertex in itself does nothing to improve the pole 
structure of the heavy quark propagator, either for the Landau gauge 
gluon propagator or the Feynman-like gluon propagator.  In particular 
we find that, as the gaussian width $\alpha$ increases, a mass pole 
pole moves in along the imaginary $k_4$ axis.  A pole on the imaginary 
axis indicates that the fermion can propagate as a free particle, and 
the position on the positive imaginary axis gives the contribution to 
the quark mass from the dynamical self dressing.  For the bare vertex, 
the pole splits into conjugate pairs either side of the imaginary axis 
as $\alpha$ decreases.  In this instance the quark becomes a confined 
particle.  Numerical difficulties described above prevented 
us from confirming that the same situation occurs in the case of the 
Ball-Chiu vertex.  As $\alpha \rightarrow 0$ we must recover the 
solution Eq.~(\ref{bcsoln}), which is an entire function with an essential 
singularity at infinity.  

We see from Table~\ref{tab} that, when poles occur, their position remains 
almost unchanged in going from Landau to Feynman-like gauge if the Ball-Chiu 
vertex is used, but not if the bare vertex is used.  In a properly 
formulated gauge covariant calculation, the position of any propagator 
mass pole should be independent of the gauge fixing procedure~\cite{AF79}.  
While we certainly do not claim that that our treatment is gauge covariant, 
it is amusing to note that replacing the bare vertex by the Ball-Chiu 
vertex (and Feynman gauge by the computationally convenient Feynman-like 
gauge) appears to go some way towards satisfying this requirement.  

\subsection{Frank and Roberts gluon propagator} 

We now return to the renormalised DSE Eq.~(\ref{hsde2}) with the gluon 
propagator $\Delta$ set equal to the smeared FR Ansatz
Eq.~(\ref{GFRprop}).  For numerical simplicity we shall restrict ourselves 
to the rainbow approximation $\Gamma_\mu(p,q) = \gamma_\mu$.  We set 
\begin{equation}
K = X + iY, \hspace{10 mm} \kappa = i\eta.  
\end{equation}
Assuming the contour of integration can be deformed to pass through 
$k'_4 = K$ for the integral of the first term in chain brackets in 
Eq.(\ref{hsde2}), and through $k'_4 = \kappa$ for the second term, we 
further set 
\begin{equation}
k'_4 = x + iY \hspace{3 mm}\mbox{ and } \hspace{3 mm} k'_4 = x + i\eta, 
\end{equation}
respectively in each of these two terms.  We also make the replacement 
$k'^2/(2m_R) \rightarrow (x^2 + \left|{\bf k}'\right|^2)/(2m_R)$ without 
affecting $\Sigma$ to leading order.  This gives 
\begin{eqnarray}
\lefteqn{ \Sigma(X + iY, i\eta) = \frac{4}{3} 
             \int_{-\infty}^\infty \frac{dx}{2\pi} 
       \int\frac{d^3{\bf k}'}{(2\pi)^3}} \nonumber \\
& &  
 \left\{  \frac{1}{ix - Y + (x^2 + \left|{\bf k}'\right|^2)/(2m_R) 
        + \Sigma(x + iY)} 
  \frac{\left|{\bf k}'\right|^2}{(x - X)^2 +  \left|{\bf k}'\right|^2}
 \Delta\left[(x - X)^2 +  \left|{\bf k}'\right|^2\right] \right. \nonumber \\
& & - \left. \frac{1}{ix - \eta + (x^2 + \left|{\bf k}'\right|^2)/(2m_R) 
                 + \Sigma(x + i\eta)}
     \, \frac{\left|{\bf k}'\right|^2}{x^2 +  \left|{\bf k}'\right|^2}
         \Delta\left[x^2 +  \left|{\bf k}'\right|^2\right]\right\} . 
                     \label{sg2}   
\end{eqnarray}

For the purpose of carrying out the numerics, it is convenient 
to change to the polar coordinates $x = r \cos\phi$, 
$\left|{\bf k}'\right| = r \sin\phi$, giving finally 
\begin{eqnarray}
\lefteqn{ \Sigma(X + iY, i\eta) = \frac{1}{3\pi^3} 
             \int_0^\infty dr \int_0^\pi d\phi\, r^3\sin^2\phi} \nonumber \\
& &  
 \left\{  \frac{1}{ir\cos\phi - Y + r^2/(2m_R) 
  + \Sigma(r\cos\phi + iY)} \frac{r^2\sin^2\phi}{r^2 - 2Xr\cos\phi +X^2}
         \Delta(r^2 - 2Xr\cos\phi +X^2) \right. \nonumber \\
& & - \left. \frac{1}{ir\cos\phi - \eta + r^2/(2m_R) 
         + \Sigma(r\cos\phi + i\eta)}\sin^2\phi \Delta(r^2)\right\}.  
                                        \label{hsde3}
\end{eqnarray}
This equation is first solved numerically along the line ${\rm Im}\,K= \eta$, 
(i.e. $Y = \eta$), and the function $\Sigma$ along this line is stored for 
subsequent calculations at arbitrary $Y$.  

\subsubsection{Numerical results: Frank and Roberts propagator}

We have numerically solved Eq.~(\ref{hsde3}) over a region of the 
complex $K$ plane with the smeared FR gluon propagator $\Delta_{\rm GFR}$.  
Our parameter choices are $m_t = 0.69$ GeV, 
in agreement with ref.~\cite{FR96}, and $\alpha 
= 16m_t^2 \hat{\alpha}d/3 = 0.5643$ (GeV)$^2$ 
corresponding to $\hat{\alpha} = 0.5$.  The choice of $\alpha$ is 
designed so that a comparison can be made between the full FR propagator 
and the gaussian propagator Eq.~(\ref{gauss}) obtained by 
keeping only the first term in Eq.~(\ref{GFRprop}).  From Table~\ref{tab} we 
know that gaussian propagator $\Delta_{\rm G}$ with the parameter choice 
$\hat{\alpha} = 0.5$ results in a pair of conjugate poles in the heavy 
quark propagator.  The heavy quark mass is set to $m_R = 5.0$ GeV.  

In Fig.~\ref{fig4} we plot the modulus $\left|\sigma_Q(K)\right|$ of the 
heavy quark propagator for the full gaussian FR propagator 
$\Delta_{\rm GFR}$ and in Fig.~\ref{fig5} plot the same quantity using 
only the gaussian part of $\Delta_{\rm G}$ with the parameters otherwise 
unchanged.  Both calculations have been done using the renormalisation point 
$\kappa = -1.0i$ GeV, and to clarify the comparison the same 
region of the $K$ plane is displayed in both plots.  
The calculation of $\sigma_Q(K)$ involves a shift of integration path 
to a contour parallel to the real axis passing through the point $K$.  
Deforming the contour to include points behind the pole 
is a numerically tedious exercise which is unlikely to enhance our 
understanding, so no results are given for the part of the $K$ plane 
in Fig.~\ref{fig5}.  

In the process of carrying out our computations, we have observed that 
the shift property resulting from changes of renormalisation point, namely 
Eq.~(\ref{shift}), is indeed respected by our numerical solutions.  In
fact, the full plot in Fig.~\ref{fig4} was pieced together by altering 
the renormalisation point to obtain solutions in strips of the complex plane 
parallel to the real axis, and using Eq.~(\ref{delm}) to match solutions 
where strips overlapped.  

We note a clear movement of the propagator pole 
further away from the real $K$ axis when the asymptotic ultraviolet tail 
is included in the gluon propagator.  Since it is the proximity of the 
pole to the real axis which prevented a solution to BSE in our earlier 
studies, this suggests that this movement of the pole portends well for 
future possible studies of heavy quark mesons if careful attention 
is paid to the asymptotic ultraviolet behaviour of the gluon propagator.  

However, a note of caution is in order.   A  different choice of 
renormalisation point would result in the plots in Figs.~\ref{fig4} 
and \ref{fig5} shifting  by different amounts respectively along the 
imaginary axis, ensuring that both plots pass through 
the same point $\sigma_Q(\kappa) = 1/i\kappa$.  Consequently, the actual 
amount by which the pole moves away from the real $K$ axis as a result 
of adding an asymptotic tail to the gluon propagator is an 
artefact of the choice of renormalisation point, though the movement will 
always be {\em away} from the real axis.  Of course only by carrying 
though the BSE calculation completely can one say for certain whether 
bound state meson solutions can be obtained.  

\section{Conclusions and outlook}

We have explored the analytic structure of heavy quark 
propagators following a recently proposed formalism which 
borrows ideas both from the DSE technique and HQET.  It is our belief that, 
if the successes of HQET are to be properly understood, we must first 
understand how the non-perturbative dynamics of QCD affect the heavy quark 
propagator.  Within the light quark sector the analytic structure of the 
quark propagator is perhaps best understood in terms of model 
Dyson-Schwinger equations.  It is therefore a worthwhile exercise to 
extend the DSE technique to the heavy quark limit.  

The initial attempt in this direction~\cite{BL97} failed essentially 
because the approximations used led to spurious propagator poles which 
prevented solution of the bound state Bethe-Salpeter equations.  Two 
approximations were involved: modelling of the quark-gluon vertex and 
of the the gluon propagator.  We have focused on each of these aspects in 
turn in this paper.  
In order to deal with an improved Ansatz for the gluon propagator with 
a realistic asymptotic ultraviolet behaviour, it has been necessary to 
formulate a properly renormalised version of the heavy quark DSE technique 
proposed in ref.~\cite{BL97}.  As an interesting corollary to our formalism 
we observe that the freedom to choose the renormalisation point is 
tantamount to the freedom in zeroth order HQET to choose the heavy quark 
mass up to a residual mass.  

We have first examined the effect of replacing the bare vertex with 
Ans\"{a}tze based on the Ball-Chiu vertex~\cite{BC80}, which is primarily 
designed to satisfy the Ward-Takahashi identity.  Specifically, we have 
considered the minimal Ball-Chiu vertex and two variants: that proposed 
by Curtis and Pennington~\cite{CP90} and that proposed by 
Haeri~\cite{H91}.  The two variants differ from the minimal vertex by the 
inclusion of extra transverse components.  
We find that, to zeroth order in the inverse of the heavy quark mass, 
the heavy quark propagator is insensitive to which of the three above 
Ans\"{a}tze is used.  One is led to question to what extent transverse 
additions to the minimal Ball-Chiu vertex can be ignored in determining 
the leading order heavy quark propagator.  If they can be be ignored in 
general there are immediate benefits in using the heavy fermion limit as 
a test-case for studies of confining field theories.  

In our numerical calculations we began with the Landau gauge form of 
the model gaussian gluon propagator $\Delta_{G}$, Eq.~(\ref{gauss}), 
which was employed (together with the bare vertex) in previous 
studies~\cite{BL97}.  Unfortunately, we find no improvement in the 
propagator pole structure in going from the bare vertex to the Ball-Chiu 
vertex while maintaining a gaussian gluon propagator.  That is to say, 
timelike mass poles indicating non-confinement, 
or conjugate poles which are likely to interfere with the successful 
solution to bound state problems, are not removed simply by improving the 
quark gluon vertex Ansatz alone.  However, in the limit in which the width 
of the gaussian gluon propagator is taken to zero (the `infrared 
dominant model'), we do obtain an entire function heavy quark propagator, 
free from singularities except an essential singularity at infinity.  
This is consistent with the equivalent finite quark mass 
calculation~\cite{BRW92}, and may provide a useful propagator for 
phenomenological modelling purposes.  

We conclude then that it is most likely the remaining approximation, 
namely the gaussian model gluon propagator, which is responsible for the 
poor analytic structure previously obtained for the quark propagator.  
To explore 
this possibility, we have replaced the simple gaussian gluon propagator 
Ansatz of ref.~\cite{BL97} by a gaussian smeared version of the more 
sophisticated Frank and Roberts Ansatz $\Delta_{\rm FR}$
given by Eq.~(\ref{FRprop}).  In this case, 
convergence of the integral in the DSE which is lost by naively retaining 
only the lowest order of the $1/m_R$ expansion of the quark propagator 
must be restored by judiciously including at least the 
spatially dependent $O(1/m_R)$ part: 
\begin{equation}
S(p) = \frac{1 + \gamma_4}{2} \frac{1}
  {ik_4 + \left|{\bf k}\right|^2/2m_R + \Sigma(k_4)} + 
                  O\left(m_R^{-1}\right).   
\end{equation}
The renormalised current quark mass $m_R$ then becomes an 
ultraviolet regulator, and in the case of the FR propagator the lowest 
order contribution to the mass expansion of the heavy quark self energy 
behaves as $\ln(m_R/m_t)$, where $m_t$ delineates the scale at which 
the asymptotic ultraviolet behaviour of the FR propagator sets in.  

Our numerical solutions of the heavy quark DSE show a clear movement of 
the offensive propagator poles away from that part of the complex momentum 
plane likely to be sampled by a Bethe-Salpeter calculation of heavy 
quark meson states.  However we caution that the amount by which the 
poles shift is, strictly speaking, dependent on the choice of renormalisation
point.  Without carrying through the the Bethe-Salpeter analysis one 
cannot say for sure that the problem is solved.  A further study of the 
BSE for heavy mesons is expected to be the focus of future work.  

Our work has also thrown up a couple of other interesting questions worthy 
of attention.  Firstly, it should be possible to check 
directly to what extent the propagator pole structure is invariant with 
respect to the choice of gauge fixing parameter $\xi$.  As the postions of 
propagator poles should be gauge independent~\cite{AF79}, this provides 
a straightforward measure of the ability of a particular vertex Ansatz 
to respect gauge covariance of the model.  

Secondly, one is led to question the meaning of propagator poles obtained 
from a Euclidean DSE formalism.  Whether propagator poles obtained in this 
way are an artefact of the approximations used or whether they are a 
genuine property of quark propagators 
has been an open question for some time~\cite{RW94,M94}\footnote{In an 
interesting recent development, McKay and Munczek~\cite{MM97} have 
examined the analytic structure of quark propagators when an extra 
constraint that solutions of the DSE should be Fourier transformable is 
imposed.  }.  
It is possible that the heavy fermion limit may help to shed some light 
on this problem by developing a Bethe-Salpeter formalism for heavy 
quarkonium states.  It is well known that the non-relativistic limit 
of the Bethe-Salpeter equation for a heavy fermion--heavy antifermion 
bound state can be written in the form of a Schr\"odinger 
equation~\cite{FS82}.  The derivation typically assumes physical mass 
poles in the fermion propagators whose residues contribute to the 
resulting Schr\"odinger 
equation.  An analogous derivation for the case of propagator poles 
which have moved off the timelike momentum axis as a result of a confining 
gluon propagator may help both with interpretation of quark propagator 
poles and with understanding the success of heavy quark potential models.  

\section*{Acknowledgement}

The author is grateful to C.\ D.\ Roberts and P.\ Maris for helpful 
discussions.  

\renewcommand{\theequation}{\Alph{subsection}.\arabic{equation}}
\setcounter{subsection}{1}
\setcounter{equation}{0}
\section*{Appendix: convergence of the heavy quark DSE}

Consider the heavy quark DSE Eq.~(\ref{hsde2}), and suppose we assume 
for the gluon propagator Ansatz the asymptotic ultraviolet behaviour 
\begin{equation}
\Delta(k^2) \sim \frac{4\pi^2 d}{k^2} \hspace{5 mm}\mbox{ for $k^2 > m_t^2$}, 
\end{equation}
where $m_t$ is a scale parameter typically of the order of 1~GeV.  
This is the behaviour exhibited by the smeared FR Ansatz 
Eq.~(\ref{GFRprop}).  We demonstrate here that the right hand side of 
Eq.~(\ref{hsde2}) is the finite difference of two logarithmically divergent 
integrals.  

We begin with the change of variables 
\begin{equation}
k'_4 = r \cos\phi, \hspace{5 mm} \left|{\bf k}'\right| = r \sin\phi, 
\end{equation}
and hence 
\begin{equation}
\int^\Lambda d^4k' = 4\pi\int dk'_4 \int d\!\left|{\bf k}'\right|
        \,\left|{\bf k}'\right|^2 
   = 4\pi\int_0^\Lambda dr \int_0^\pi d\phi \,r^3 \sin^2\phi.  
\end{equation}
This gives 
\begin{eqnarray}
\Sigma(K,\kappa) & = & \frac{1}{3\pi^3} \int_0^\Lambda dr \int_0^\pi d\phi\,
  \frac{r^5\sin^4\phi }{ir\cos \phi + r^2/(2m_R) + \Sigma(r\cos\phi)} 
                     \nonumber \\
 & & \times \left[ \frac{\Delta(K^2 - 2Kr\cos\phi + r^2)}
  {K^2 - 2Kr\cos\phi + r^2} - (K\rightarrow\kappa) \right] .  \label{pol}
\end{eqnarray}
Implicit in this equation is a hierarchy of scales given by Eq.~(\ref{hier}).  
Simply counting powers of $r$ in the integrand, we see that each of the 
two terms diverges as 
\begin{equation}
m_R\int^\Lambda \frac{dr}{r} \sim m_R \ln \Lambda.  
\end{equation}

On the other hand consider the difference of the two terms.  For $r > m_t$ 
the part in square brackets is 
\begin{eqnarray}
\lefteqn{ \left[ \frac{4\pi^2 d}
            {(K^2 - 2Kr\cos\phi + r^2)^2} - (K\rightarrow\kappa) \right] }
                   \nonumber \\
& = & 4\pi^2d\left[\frac{4(K - \kappa)\cos\phi}{r^5} + 
     \frac{2(K^2 - \kappa^2)(6\cos^2\phi - 1)}{r^6}\right] + 
             O\left(\frac{1}{r^7}\right).  
\end{eqnarray}
Neglecting $\Sigma(r\cos\phi)$ in Eq.~(\ref{pol}) for large $r$, we can 
approximate the propagator contribution to the integrand by 
\begin{equation}
\frac{1}{ir \cos\phi + r^2/(2m_R)} = 
         - \frac{i\cos\phi - r/2m_R}{r(\cos^2\phi + r^2/4m_R^2)}.  
\end{equation}
Then, taking into account the hierarchy (\ref{hier}), the contribution 
to the integrand for $r>m_t$ is approximately 
\begin{equation}
\frac{16id(\kappa - K)}{3\pi}\int_{m_t}^\infty \frac{dr}{r} 
\int_0^\pi d\phi\,\frac{\sin^4\phi \cos^2\phi}{\cos^2\phi + r^2/4m_R^2}
\end{equation}
giving
\begin{equation}
 \Sigma(K,\kappa) \sim  2id(\kappa - K) \ln\left(\frac{m_R}{m_t}\right) 
              \hspace{5 mm}\mbox{as $m_R\rightarrow\infty$}.  
\end{equation}
The last step can be achieved using the crude approximation 
\begin{equation}
\frac{\cos^2\phi}{\cos^2\phi + r^2/4m_R^2} \approx \left\{ 
   \begin{array}{lr} 1                         & \mbox{ if $r<4m_R$} \\
                     (4m_R^2/r^2)\cos^2\phi    & \mbox{ if $r>4m_R$} 
                         \end{array} \right.    .
\end{equation}
%

%
\pagebreak
\begin{table} 
\caption{Position $k_4/\mu$ of poles in the heavy quark propagator
  closest to the real $k_4$ axis, using the gaussian gluon propagator 
  Ansatz $\Delta_{\rm G}$.
  Numerical instabilities prevent an accurate location of the pole 
  in the case indicated by a question mark.}
                             \label{tab}
\begin{center}
\begin{tabular}{|ccccc|}\hline
            &  \multicolumn{2}{c}{Landau gauge} &
              \multicolumn{2}{c|}{Feynman-like gauge} \\ \hline
$\hat{\alpha} = \alpha/\mu^2$ 
            & bare vertex & BC vertex & bare vertex & BC vertex  \\ \hline 
   0.5  &$\pm0.510 + 0.415i$ &    ?     &$\pm0.5466 + 0.5109i$&     ?     \\
    1   &$\pm0.378 + 0.506i$ &    ?     &$\pm0.3383 + 0.6758i$& $0.4429i$ \\
    2     &     $0.466i$     &  $0.31i$ &      $0.3844i$      & $0.3130i$ \\
    3     &     $0.307i$     & $0.252i$ &      $0.2880i$      & $0.2554i$ \\
    4     &     $0.251i$     & $0.219i$ &      $0.2408i$      & $0.2210i$ \\
\end{tabular}
\end{center}
\end{table}
%
%
\pagebreak
\begin{figure}
\caption{The change of variables $p^2 \rightarrow K$ used to represent 
the heavy quark propagator in the region of the bare fermion mass pole
$p^2 = -m_R^2$.  For the change of variables in the DSE to be valid, the 
quark propagator must be analytic over the shaded region.   
\label{fig1}}
\end{figure}
\begin{figure}
\caption{The heavy quark self energy $\Sigma(k_4)$, 
from the DSE in Landau gauge with a bare quark-gluon vertex and gaussian
gluon propagator with parameters: $\mu = 1$ and $\alpha = 0$ (solid curve), 
1 (long dashes), 2 (short dashes) and 3 (dotted curve).  
The upper curves are ${\rm Re}\,\Sigma$ and the lower curves 
${\rm Im}\,\Sigma$.  
\label{fig2}}
\end{figure}
\begin{figure}
\caption{The same as Fig.~\ref{fig2}, except with the Ball-Chiu 
quark-gluon vertex.  
\label{fig3}}
\end{figure}
\begin{figure}
\caption{The modulus $\left|\sigma_Q(K)\right|$ of the heavy quark 
propagator in the complex $K$ plane obtained by solving the heavy quark 
DSE in rainbow approximation with a gaussian smeared Frank and Roberts 
Ansatz $\Delta_{\rm GFR}$ for the gluon propagator.  Input parameter values 
are given in the text. 
\label{fig4}}
\end{figure}
\begin{figure}
\caption{The same as Fig.~\ref{fig4}, except with the gluon propagator 
Ansatz replaced by only its Gaussian part $\Delta_{\rm G}$, and all 
parameter values otherwise unchanged.  
\label{fig5}}
\end{figure}
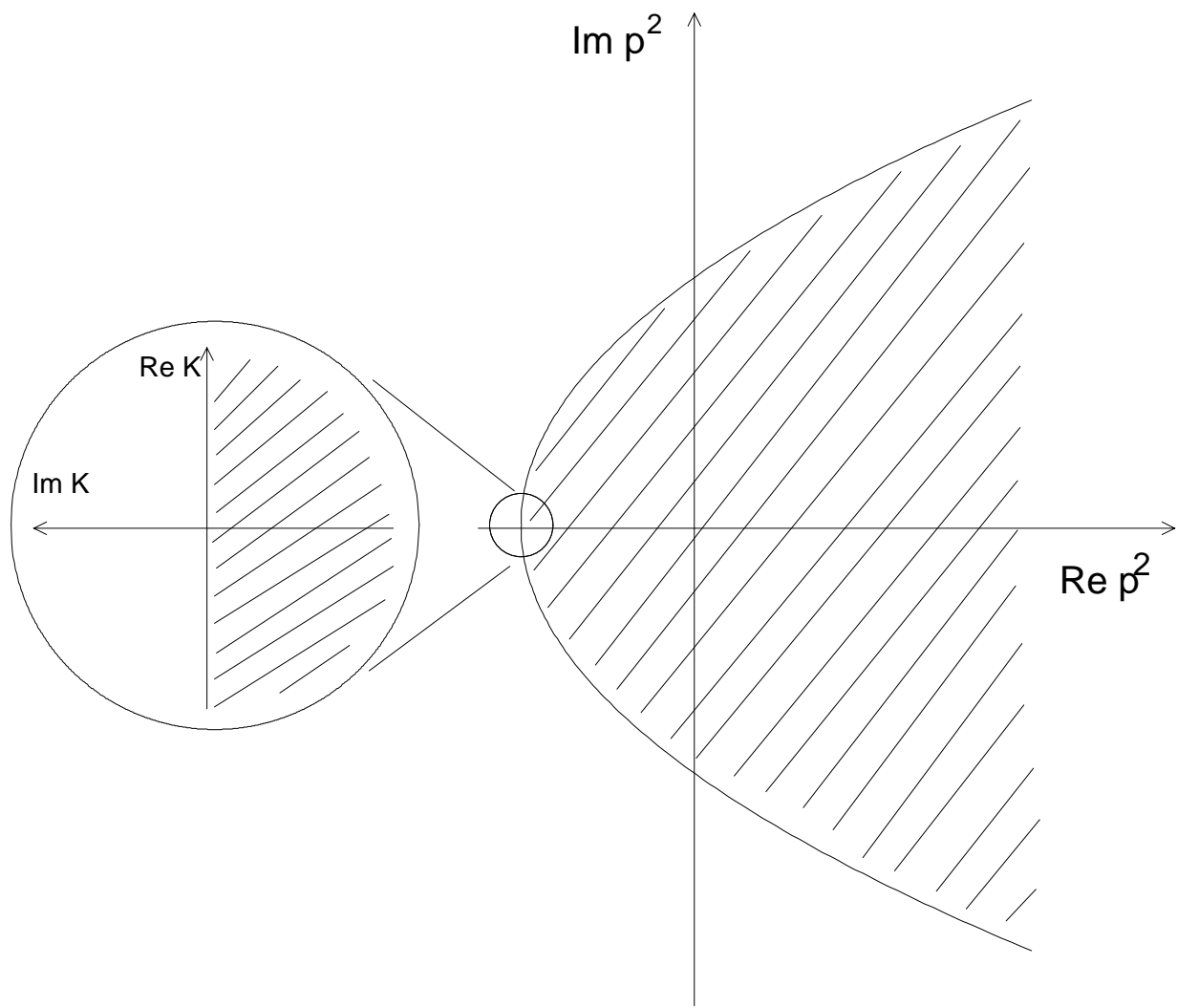
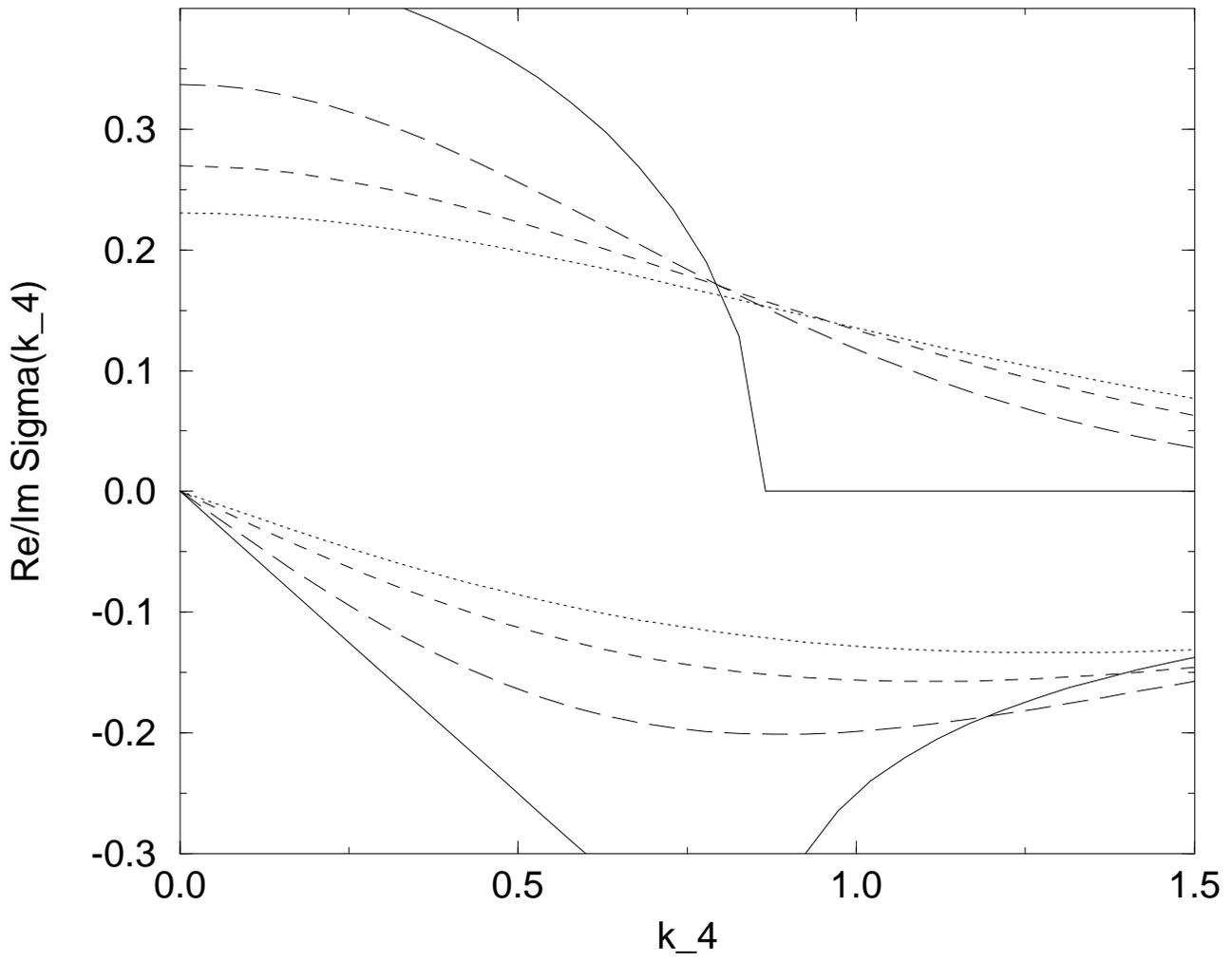
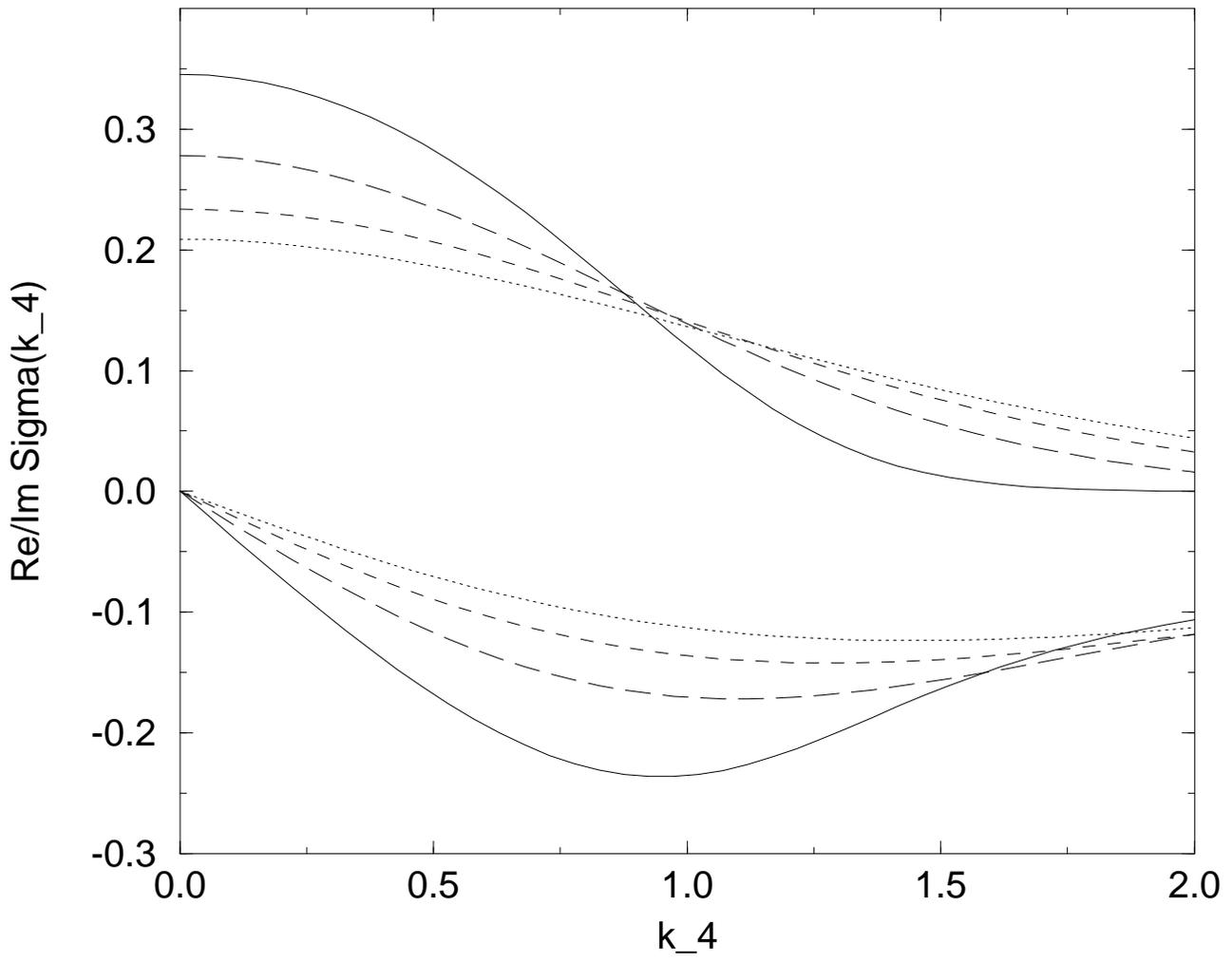
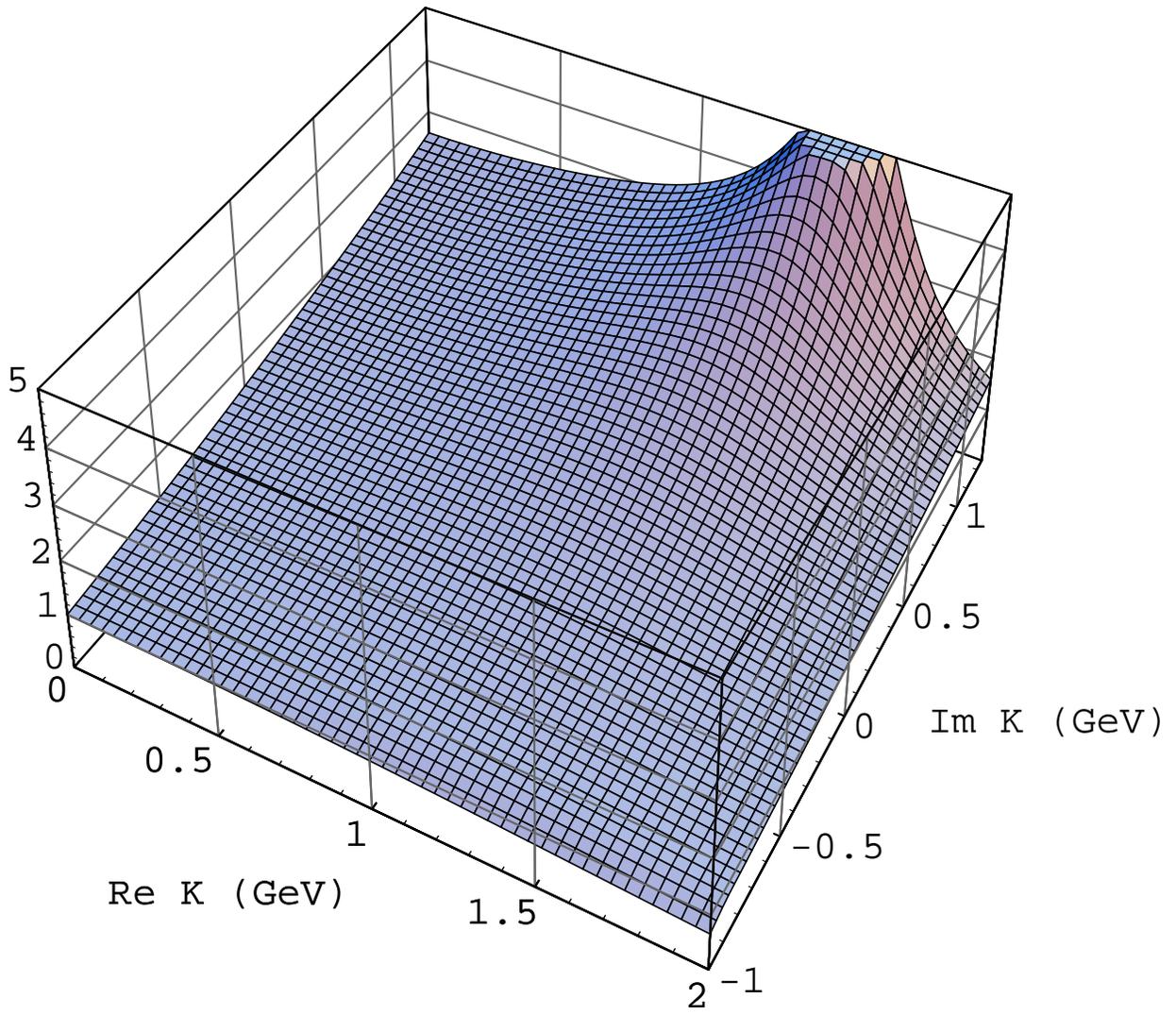
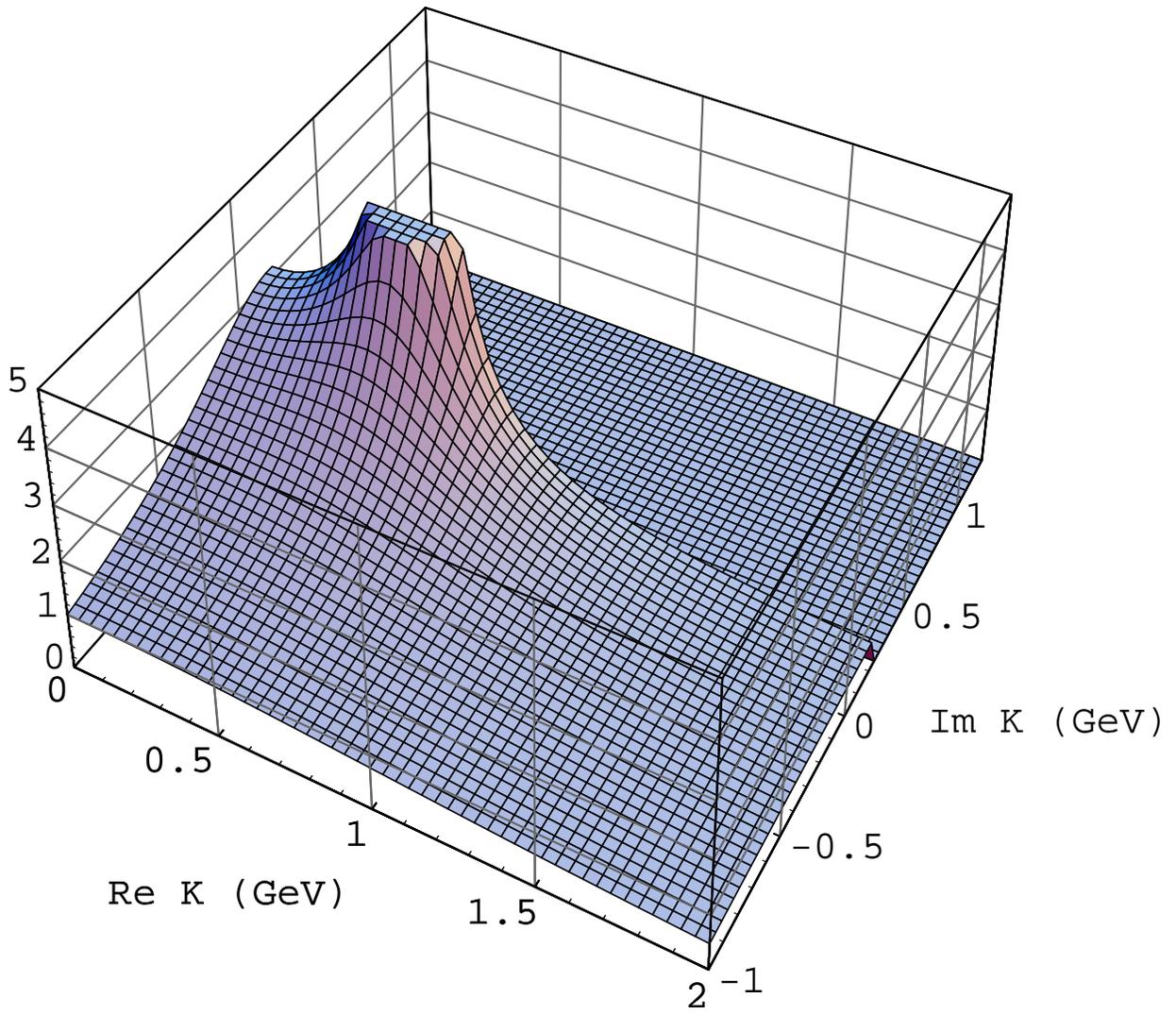
\end{document}